\documentclass[11pt]{article}
\usepackage{blois,epsfig}
\usepackage[latin1]{inputenc}
\usepackage[OT1]{fontenc}

\def\be{\begin{equation}}
\def\ee{\end{equation}}
\def\bea{\begin{eqnarray}}
\def\eea{\end{eqnarray}}

\def\av{{\sc Antares/Virgo}}
\def\anta{{\sc Antares}}
\def\vo{{\sc Virgo}}
\def\lo{{\sc Ligo}}
\begin{document}

\vspace*{4cm}
\title{{\sc Antares/Virgo} Coincidences : a feasibility study}

\author{Thierry PRADIER}

\address{{\sl Institut Pluridisciplinaire Hubert Curien} {\sc (iphc/drs)} \\ 
\& University Louis-Pasteur, Strasbourg (France)}

\maketitle\abstracts{Sources of gravitational waves (GW) and emitters of high energy (HE) neutrinos both involve compact objects and matter moving at relativistic speeds. Coincidences between \vo~and \anta~would give a unique insight on the physics of the most powerful objects in the Universe. 
The feasibility of such GW/HE $\nu$ coincidences is analysed.
}

\noindent
The forthcoming years should be very exciting both in GW astronomy and HE $\nu$ astronomy. The \vo~interferometer~\cite{virgo}, currently closed down for upgrade, should be taking 
data with an improved injection system in 2009, whereas the \anta~collaboration has completed the deployment 
and connections of its 12 lines~\cite{antares} early in June 2008, starting the operation of the first {\it undersea} 
neutrino telescope. The two \lo~interferometers (ITF) are in operation~\cite{ligo}, and {\sc IceCube} 
is deploying its 1~km$^3$ neutrino telescope in the ice of the South Pole, already having 40 lines in data taking 
mode \cite{icecube}. 
Both GW sources and HE $\nu$ emitters involve compact objects and matter moving at relativistic speeds. As a result, time coincidences between \vo~and \anta~can be envisaged. Some astrophysical objects, invisible in electromagnetic
 channels, may be observable only {\it via} their GW and HE $\nu$ emissions. 
Finally, in many quantum gravity (QG) models~\cite{qg}, the propagation velocity of a particle depends on the 
energy:
measuring a non-zero time delay between GW bursts and HE $\nu$ transients would allow to probe QG effects.

\section{GW bursters and HE $\nu$ sources : the case for microquasars major outbursts}
\label{sec:sources}

\begin{figure}[h!]
\centerline{\includegraphics[width=0.4\linewidth]{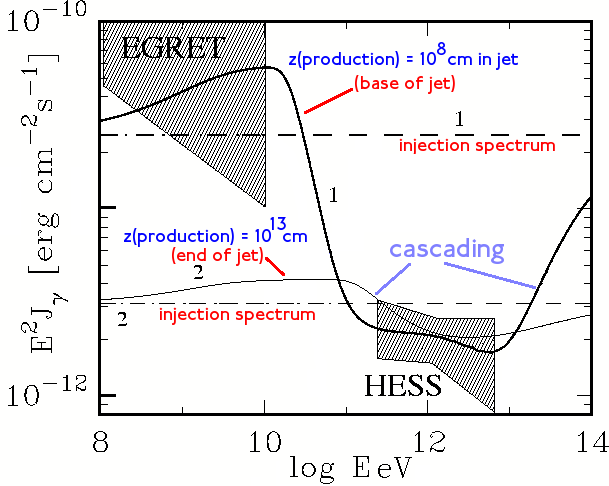}\includegraphics[width=0.327\linewidth]{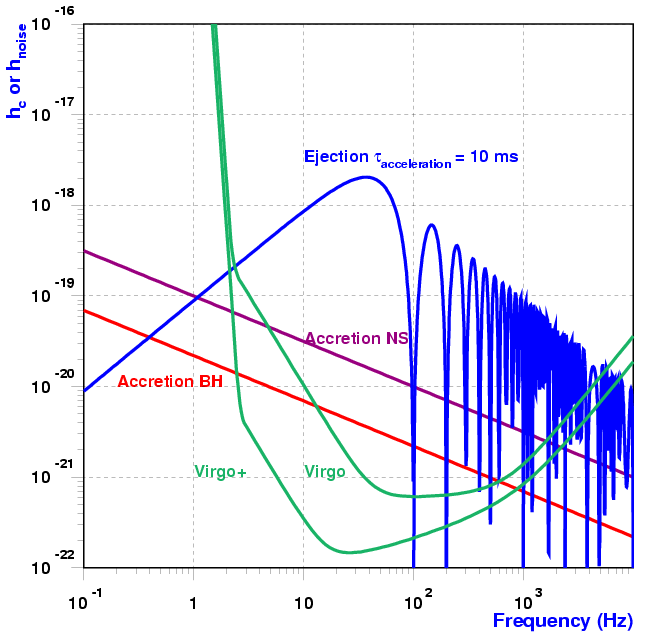}}
\caption{Left : $\gamma$ Data from EGRET and HESS, together with predictions for models with production of $\gamma$ at the base or at the end of the jet. Right : Amplitudes expected for the accretion/ejection processes at 1~kpc, for $\delta m \sim 10^{-4} \textrm{M}_{\odot}$, $\tau_{\textrm{acc}} \sim 10$~ms, $\gamma = 10$.
\label{fig:microq}}
\end{figure}

Flares from soft-gamma repeaters~\cite{sgr}, core-collapse supernovae and gamma-ray bursts are commonly cited sources both for GW bursts and neutrino emissions, but are described elsewhere~\cite{toulon}. 
Microquasars are galactic jet sources associated with
some classes of X-ray binaries involving both neutron
stars and black hole candidates~\cite{mirabel}. 
The content of jets in microquasars remains an open issue.
In scenarios in which an initial rise
of the X-ray flux leads to ejection of the inner part of the
accretion disk, as is widely claimed to be suggested by the
anticorrelation between the X-ray and radio flares seen during
major ejection events,
e-p jets are expected to be produced. A
possible diagnostic of e-p jets is the presence of Doppler-shifted
spectral lines, such as the H${\alpha}$ line as seen in SS433.
Taking the example of LS 5039, some authors argue in favor of hadronic origin of TeV photons~\cite{ls5039}.
The
flux of TeV neutrinos estimated on the basis of the detected TeV 
$\gamma$-ray flux depends significantly on the location of $\gamma$-ray
production region. 
The {\sc Hess/Egret} data agree well with a production of $\gamma$/$\nu$ at the base of the jet, as seen in figure~\ref{fig:microq}~\cite{ls5039}, close to the onset of the acceleration phase and its corresponding GW signal.

The matter accreted for months/years could be {\it swallowed} by the compact object~\cite{toulon}, and, provided that the process is fast, trigger the resonance of normal modes of the central object, typically a damped sine signal, which could last during the ejection phase. The acceleration of an ultrarelativistic blob of matter with a Lorentz factor $\gamma$ around a compact object induces a {\it burst with memory} independent on the density of the ejecta~\cite{jetgw}. The frequency is typically the inverse of the acceleration time $t_{\textrm{{\tiny acc}}}$. For both signals, the amplitude will critically depend on the accreted/ejected mass, which can be estimated to range from $10^{-8} \textrm{M}_{\odot}$ up to $10^{-4} \textrm{M}_{\odot}$, for major events.
A summary of these estimates for the most extreme outbursts are displayed in Figure~\ref{fig:microq}.

\section{Observability of \anta/\vo~Coincidences}
\label{sec:obs}

\begin{figure}[h!]
\centerline{\includegraphics[width=0.32\linewidth]{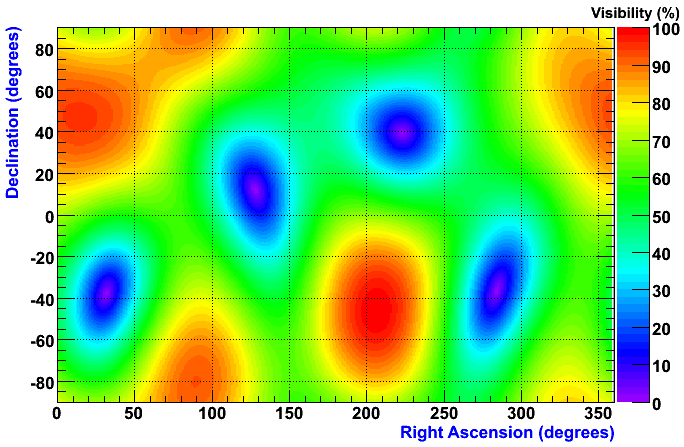}\includegraphics[width=0.34\linewidth]{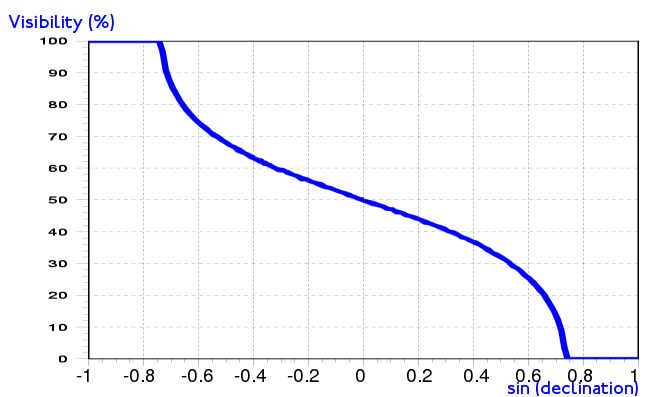}\includegraphics[width=0.34\linewidth]{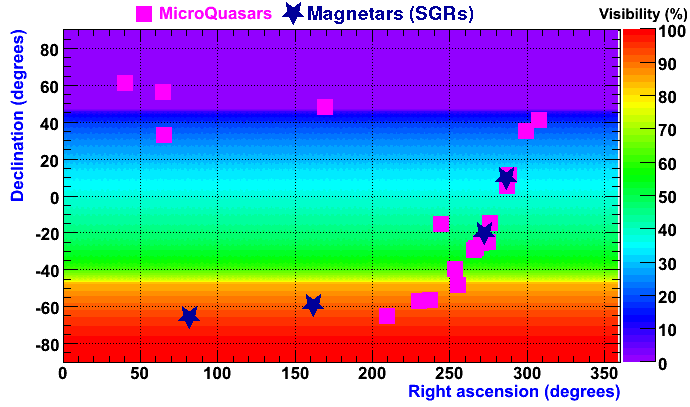}}
\caption{Left : \vo~beam pattern in equatorial coordinates. Middle : \anta~daily averaged visibility {\it vs} $\sin \delta$. Right : Common visibility sky map for \anta~and \vo.
\label{fig:visu}}
\end{figure}

Both \vo~and \anta~have limited sky coverage and exposure. In order to perform coincidences between both detectors, the overlap of such visibility maps has to be computed.
The response $h(t)$ of an interferometric detector to a GW is a linear combination of the two independent wave polarizations $h_+$ and $h_{\times}$, with weighting factors called the beam pattern functions.
 The instantaneous beam pattern (normalised to its maximal value, and averaged over the unknown polarization angle) in equatorial coordinates (right ascension $\alpha$, declination $\delta$) is displayed in figure~\ref{fig:visu} for \vo. 

Searching for neutrinos 
which have interacted in the Earth, \anta~is only sensitive to sources below the horizon at some time during the day. Figure~\ref{fig:visu} 
shows the daily average visibility as a function of $\sin \delta$.
Finally, the visibility sky map for coincidences between \anta~and \vo~is the convolution of the two previous exposure maps. The daily averaged common sky map is displayed in figure~\ref{fig:visu}, together with the position of known microquasars and soft-gamma repeaters (or magnetars). Most of these galactic sources are visible at some time by both experiments, rendering observable GW/HE $\nu$ coincidences for these sources.

\section{Detectability of \anta/\vo~Coincidences}
\label{sec:detec}

To set the coincidence time window, possible physical propagation delays have to be estimated. In the case of GW, 
the graviton being massless, and the energy carried away by each individual graviton in a GW burst being small 
($E_{\textrm{{\tiny graviton}}}\sim hf\ll 1$ for $f = 1~\textrm{kHz}$), QG-induced or mass-induced delays are 
close to zero. For a 1 TeV $\nu$, the mass-induced delay is negligible even with $m_{\nu} = 1~\textrm{eV}$. 
Taking the expression given in~\cite{qg} as a starting point, neglecting any cosmological effects (for low redshift $z \ll 1$), the delay in ms becomes, in the first order 
$\Delta t^{\textrm{{\tiny ms}}}_{\textrm{{\tiny QG}}} \propto 1/d \simeq 0.15 \left(\frac{E_{\nu}}{1~\textrm{TeV}}\right)\left(\frac{10^{19}~\textrm{GeV}}{E_{\textrm{{\tiny QG}}}}\right)$ at 10~kpc, where $E_{\textrm{{\tiny QG}}}$ is the energy scale at which QG effects arise. Taking $E_{\textrm{{\tiny QG}}} = E_{\textrm{{\tiny Planck}}} = 10^{19}~\textrm{GeV}$, this yields a maximum 
QG delay of 1 second for $E_{\nu} \approx 1~\textrm{TeV}$ and sources as far as the Virgo Cluster 
($d \sim 20~\textrm{Mpc}$). $\Delta t_{\textrm{{\tiny coinc}}} = 1~$s thus seems a reasonnable choice. 
The coincidence time window can also be set by imposing an overall coincidence detection 
probability.

\begin{figure}[h!]
\centerline{\includegraphics[width=0.6325\linewidth]{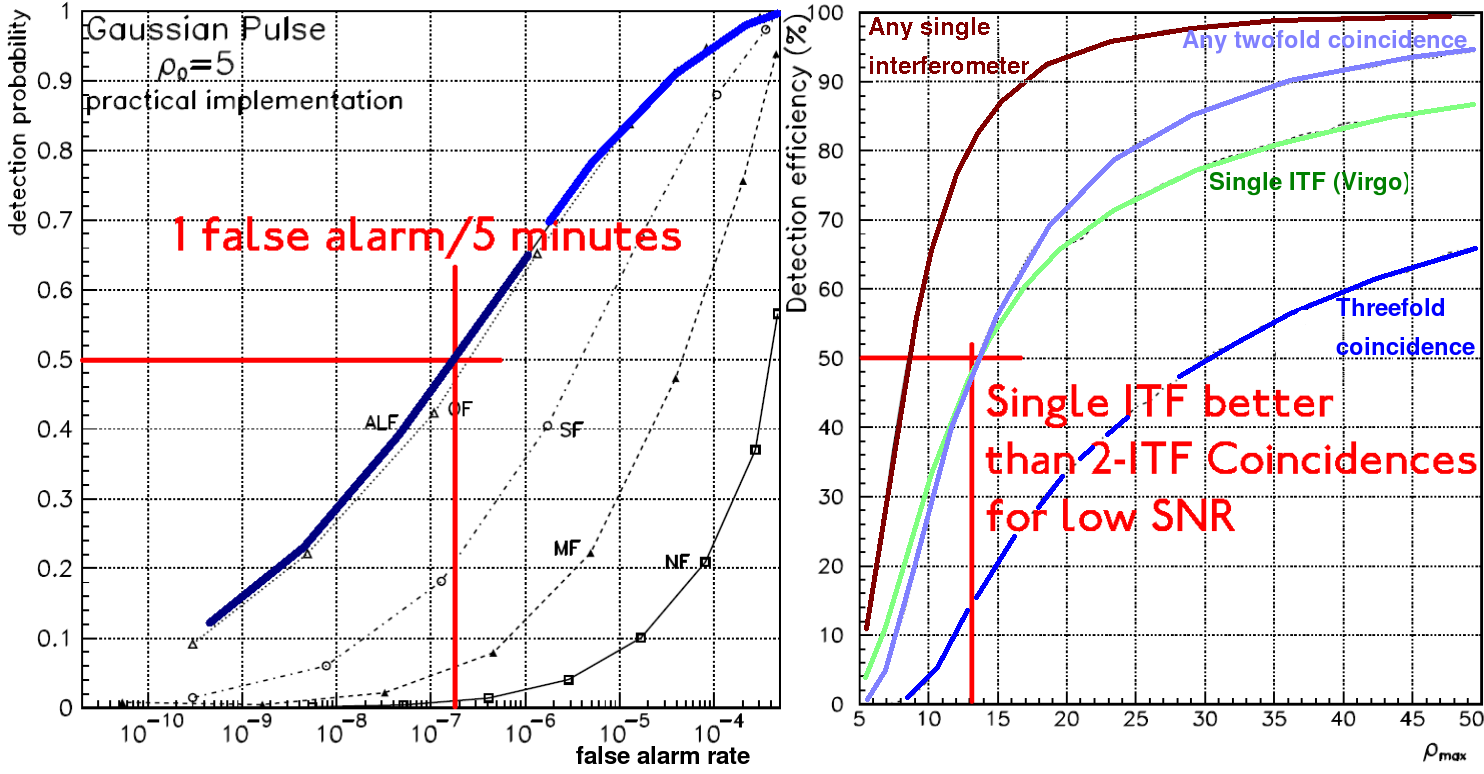}\includegraphics[width=0.3675\linewidth]{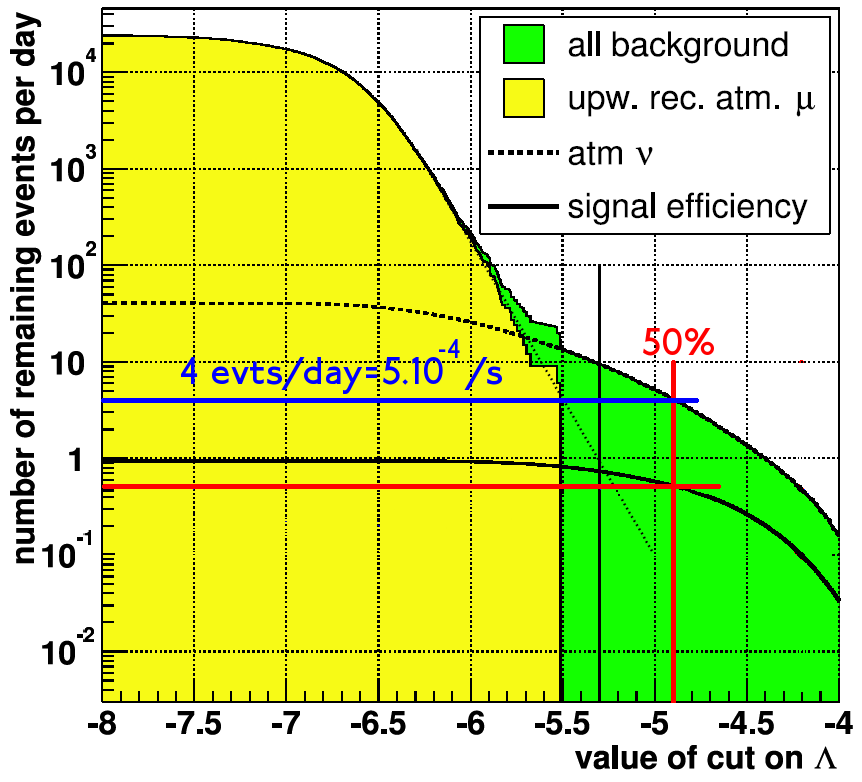}}
\caption{Left : detection probability {\it vs} false alarm rate, for a $\rho = 5$ gaussian pulse, $\Delta t_{\textrm{{\tiny coinc}}} = 1~$s. Middle : detection probability {\it vs} $\rho_{\textrm{{\tiny max}}}$ for the different possible detection configurations. Right : Number of background events left as a function of the cut value for $\Lambda$. The efficiency for signal is also shown.
\label{fig:detection}}
\end{figure}

The detection probability for \vo~can be estimated as a function of the signal-to-ratio (SNR or $\rho$) of a particular 
signal, for a given false alarm rate: this is shown for \vo~in figure~\ref{fig:detection}, for $\rho_{\textrm{{\tiny max}}} = 5$ (see~\cite{toulon,virgodetection} for details). In the best case, a threshold corresponding 
to 1 false alarm every 5 minutes is needed to obtain a 50\% detection probability (no beam pattern effects included).
The middle panel of figure~\ref{fig:detection} displays the detection efficiency as a function of $\rho_{\textrm{{\tiny max}}}$, 
for this particular false alarm rate, in the case of a single ITF detection or coincident detection in the 
\vo/\lo~network: for low SNRs, the detection by a single ITF is more probable than any twofold coincidence, 
and the detection by any single ITF is always more efficient than any coincidence configuration~\cite{virgodetection}. 
In the case of the detection by any of the 3 ITFs, the directional information is not available (no triangulation), 
and the only relevant information is therefore the {\it time} of the burst event. Finally, the rms error on the burst arrival time~\cite{virgodetection} for a gaussian burst 
of width $\tau$ and SNR $\rho$ is $\Delta t^{\textrm{{\tiny RMS}}} \approx \frac{1.5}{\rho} \left(\frac{\tau}{1~\textrm{ms}} \right)~\textrm{ms}$. This of course limits the accessible 
QG energy scale, and the coincidence window to be used.

In \anta, the \v{C}erenkov light emitted
by the neutrino-induced muon is detected by an array of photomultipliers arranged in strings, able to reconstruct the energy and 
direction of
the incident muon/neutrino. 
The quality of the track fit is often expressed in terms of a 
log-likelihood ratio term which distribution is shown in figure~\ref{fig:detection}, for (upward) atmospheric neutrinos and misreconstructed atmospheric muons, together with the signal detection efficiency for a $E^{-2}$ spectrum.
The standard cut 
applied is $\Lambda = -5.3$, for which the signal efficiency is close to 75\%~\cite{niccolo}; the atmospheric $\nu$ background is reduced to 10/day.

The plots in figure \ref{fig:detection} provide the information needed to estimate 
the detection probability $\epsilon_{{\tiny V,A}}$ for a background/false alarm level $R_{{\tiny V,A}}$ in both 
detectors (V for \vo, A for \anta). The coincidence detection probability is 
$\epsilon_{\textrm{{\tiny coinc}}} = \epsilon_{{\tiny V}} \epsilon_{{\tiny A}}$, whereas the coincident accidentals 
rate in a given time window is $R_{\textrm{{\tiny coinc}}} = R_{{\tiny V}} R_{{\tiny A}} \Delta t_{\textrm{{\tiny coinc}}}$. 
Setting $\Delta t_{\textrm{{\tiny coinc}}} = 1$~s and $R_{\textrm{{\tiny coinc}}} \sim 1/$yr, 
the resulting coincidence detection probability peaks at $\sim$ 25\%, for $\Lambda \sim -5.5$.
Equivalently, the coincidence detection probability can be set at {\it e.g.} 50\% for a given signal, and 
the maximal allowed coincidence time window can be extracted. The time coincidence window is maximal for $\epsilon_{{\tiny V}} \sim 65\%$, 
reaching $\sim 15$~ms, for a $\rho=5$ gaussian burst, with 
$R_{\textrm{{\tiny coinc}}} \sim 1/$yr~\cite{toulon}.


\section{\av~Coincidences}
\label{sec:coinc}

Assuming that 
both the GW and HE $\nu$ signals have been emitted with zero delay at the source, limits can be put on the QG energy 
scale $E_{\textrm{{\tiny QG}}}$.
Requiring $\epsilon_{\textrm{{\tiny coinc}}} = 50 \%$, the maximum energy scale yielding 
a measurable effect is limited by GW timing resolution, which depends on the burst duration and SNR, and reaches in 
this case $E^{\textrm{{\tiny max}}}_{\textrm{{\tiny QG}}} \sim 5\times10^{18}$~GeV ($\rho=5$), close to the Planck limit. 
The minimum accessible energy scale is in turn determined by the maximal coincidence window defined previously, which 
yields $E^{\textrm{{\tiny min}}}_{\textrm{{\tiny QG}}} \sim 10^{17}$~GeV. This is to be compared with existing limits 
on $E_{\textrm{{\tiny QG}}}$, {\it e.g.} using TeV flares from Mrk421 $\sim 4\times10^{17}$~GeV \cite{qgexp}. To perform a real {\it measurement} of $E_{\textrm{{\tiny QG}}}$, the neutrino energy resolution is of 
importance, and is a factor 2 or 3 in the case of \anta~\cite{antares}.

\vo~took data jointly with the 2 \lo~interferometers between May and September 2007, 
during the {\it Virgo Scientific Run} (VSR), with a final sensitivity close to the expectation. 
The interferometer should resume taking data again with an improved injection system in 2009.
\anta~has been continuously taking data with its final 12 line configuration since the end of May 2008, and is expected to observe high energy neutrinos for a period of 10 years. Interestingly, during the {\it VSR}, \anta~already had 5 lines operational since January 2007. Clearly, only the most powerful GW/HE $\nu$ sources could be detected, but
this could be used as a {\it test bench} for preliminary studies on time coincidences. \lo~data could also be used to enhance the detection probability.

Finally, {\it circa} 2015, a km$^3$ neutrino telescope should be operating in the Mediterranean Sea~\cite{km3}, along with 
an {\sc Advanced Virgo} interferometer~\cite{virgo}, with enhanced sensitity at low frequency:
less extremes accretion/ejection scenarios could be probed for microquasars, and interesting constraints on accretion/ejection models could 
be set by this novel multimessenger approach.

\end{document}